 \input miniltx
  \def\Gin@driver{pdftex.def}
  \input color.sty
  \input graphicx.sty
  \resetatcatcode
  
%
%

%
%
%
%

\def\Serif{cmr}
\def\SerifBold{cmbx}
\def\SerifItalics{cmti}
\def\SerifSlanted{cmsl}
\def\SerifBoldItalics{cmbxti}
\def\SansSerif{cmss}
\def\SansSerifBold{cmssbx}
\def\SansSerifItalics{cmssi}
\def\SansSerifSlanted{cmssi}
\def\Math{cmmi}
\def\Symbols{cmsy}
\def\MathBold{cmmib}
\def\MoreSymbols{cmex}
\def\Typewriter{cmtt}
\def\Gothic{eufm}
\def\Double{msbm}
\def\Relazioni{msam}

= 			\Serif10 			at 5pt
= 		\SerifBold10 		at 5pt
= 	\SerifItalics10 	at 5pt
=		\SerifSlanted10 	at 5pt
=	\SerifBoldItalics10	at 5pt
= 		\SansSerif10 		at 5pt
=	\SansSerifBold10	at 5pt
=	\SansSerifItalics10	at 5pt
=	\SansSerifSlanted10	at 5pt
=				\Math10				at 5pt
=			\MathBold10			at 5pt
=			\Symbols10			at 5pt
=		\MoreSymbols10		at 5pt
=		\Typewriter10		at 5pt
=			\Gothic10			at 5pt
=			\Double10			at 5pt

= 			\Serif10 			at 7pt
= 		\SerifBold10 		at 7pt
= 	\SerifItalics10 	at 7pt
=	\SerifSlanted10 	at 7pt
=\SerifBoldItalics10	at 7pt
= 		\SansSerif10 		at 7pt
= 	\SansSerifBold10 	at 7pt
=\SansSerifItalics10	at 7pt
=\SansSerifSlanted10	at 7pt
=			\Math10				at 7pt
=		\MathBold10			at 7pt
=			\Symbols10			at 7pt
=		\MoreSymbols10		at 7pt
=		\Typewriter10		at 7pt
=			\Gothic10			at 7pt
=			\Double10			at 7pt

= 			\Serif10 			at 8pt
= 		\SerifBold10 		at 8pt
= 	\SerifItalics10 	at 8pt
=	\SerifSlanted10 	at 8pt
=\SerifBoldItalics10	at 8pt
= 		\SansSerif10 		at 8pt
= 	\SansSerifBold10 	at 8pt
=\SansSerifItalics10 at 8pt
=\SansSerifSlanted10 at 8pt
=			\Math10				at 8pt
=		\MathBold10			at 8pt
=			\Symbols10			at 8pt
=		\MoreSymbols10		at 8pt
=		\Typewriter10		at 8pt
=			\Gothic10			at 8pt
=			\Double10			at 8pt

= 			\Serif10 			at 10pt
= 		\SerifBold10 		at 10pt
= 		\SerifItalics10 	at 10pt
=		\SerifSlanted10 	at 10pt
=	\SerifBoldItalics10	at 10pt
= 		\SansSerif10 		at 10pt
= 	\SansSerifBold10 	at 10pt
= 	\SansSerifItalics10 at 10pt
= 	\SansSerifSlanted10 at 10pt
=				\Math10				at 10pt
=			\MathBold10			at 10pt
=			\Symbols10			at 10pt
=		\MoreSymbols10		at 10pt
=		\Typewriter10		at 10pt
=			\Gothic10			at 10pt
=			\Double10			at 10pt
=			\Relazioni10			at 10pt

= 				\Serif10 			at 12pt
= 			\SerifBold10 		at 12pt
= 		\SerifItalics10 	at 12pt
=		\SerifSlanted10 	at 12pt
=	\SerifBoldItalics10	at 12pt
= 			\SansSerif10 		at 12pt
= 		\SansSerifBold10 	at 12pt
= 	\SansSerifItalics10 at 12pt
= 	\SansSerifSlanted10 at 12pt
=				\Math10				at 12pt
=			\MathBold10			at 12pt
=			\Symbols10			at 12pt
=		\MoreSymbols10		at 12pt
=			\Typewriter10		at 12pt
=				\Gothic10			at 12pt
=				\Double10			at 12pt

= 			\Serif10 			at 14pt
= 		\SerifBold10 		at 14pt
= 	\SerifItalics10 	at 14pt
=		\SerifSlanted10 	at 14pt
=	\SerifBoldItalics10	at 14pt
= 		\SansSerif10 		at 14pt
= 	\SansSerifBold10 	at 14pt
= \SansSerifSlanted10 at 14pt
= \SansSerifItalics10 at 14pt
=				\Math10				at 14pt
=			\MathBold10			at 14pt
=			\Symbols10			at 14pt
=		\MoreSymbols10		at 14pt
=		\Typewriter10		at 14pt
=			\Gothic10			at 14pt
=			\Double10			at 14pt

\def\NormalStyle{\parindent=5pt\parskip=3pt\normalbaselineskip=14pt%
\def\nt{\tenSerif}%
\def\rm{\fam0\tenSerif}%
\textfont0=\tenSerif\scriptfont0=\sevenSerif\scriptscriptfont0=\fiveSerif
\textfont1=\tenMath\scriptfont1=\sevenMath\scriptscriptfont1=\fiveMath
\textfont2=\tenSymbols\scriptfont2=\sevenSymbols\scriptscriptfont2=\fiveSymbols
\textfont3=\tenMoreSymbols\scriptfont3=\sevenMoreSymbols\scriptscriptfont3=\fiveMoreSymbols
\textfont\itfam=\tenSerifItalics\def\it{\fam\itfam\tenSerifItalics}%
\textfont\slfam=\tenSerifSlanted\def\sl{\fam\slfam\tenSerifSlanted}%
\textfont\ttfam=\tenTypewriter\def\tt{\fam\ttfam\tenTypewriter}%
\textfont\bffam=\tenSerifBold%
\def\bf{\fam\bffam\tenSerifBold}\scriptfont\bffam=\sevenSerifBold\scriptscriptfont\bffam=\fiveSerifBold%
\def\cal{\tenSymbols}%
\def\greekbold{\tenMathBold}%
\def\gothic{\tenGothic}%
\def\Bbb{\tenDouble}%
\def\LieFont{\tenSerifItalics}%
\def\Rel{\tenRelazioni}%
\nt\normalbaselines\baselineskip=14pt%
}

\def\TitleStyle{\parindent=0pt\parskip=0pt\normalbaselineskip=15pt%
\def\nt{\fourteenSansSerifBold}%
\def\rm{\fam0\fourteenSansSerifBold}%
\textfont0=\fourteenSansSerifBold\scriptfont0=\tenSansSerifBold\scriptscriptfont0=\eightSansSerifBold
\textfont1=\fourteenMath\scriptfont1=\tenMath\scriptscriptfont1=\eightMath
\textfont2=\fourteenSymbols\scriptfont2=\tenSymbols\scriptscriptfont2=\eightSymbols
\textfont3=\fourteenMoreSymbols\scriptfont3=\tenMoreSymbols\scriptscriptfont3=\eightMoreSymbols
\textfont\itfam=\fourteenSansSerifItalics\def\it{\fam\itfam\fourteenSansSerifItalics}%
\textfont\slfam=\fourteenSansSerifSlanted\def\sl{\fam\slfam\fourteenSerifSansSlanted}%
\textfont\ttfam=\fourteenTypewriter\def\tt{\fam\ttfam\fourteenTypewriter}%
\textfont\bffam=\fourteenSansSerif%
\def\bf{\fam\bffam\fourteenSansSerif}\scriptfont\bffam=\tenSansSerif\scriptscriptfont\bffam=\eightSansSerif%
\def\cal{\fourteenSymbols}%
\def\greekbold{\fourteenMathBold}%
\def\gothic{\fourteenGothic}%
\def\Bbb{\fourteenDouble}%
\def\LieFont{\fourteenSerifItalics}%
\nt\normalbaselines\baselineskip=15pt%
}

\def\PartStyle{\parindent=0pt\parskip=0pt\normalbaselineskip=15pt%
\def\nt{\fourteenSansSerifBold}%
\def\rm{\fam0\fourteenSansSerifBold}%
\textfont0=\fourteenSansSerifBold\scriptfont0=\tenSansSerifBold\scriptscriptfont0=\eightSansSerifBold
\textfont1=\fourteenMath\scriptfont1=\tenMath\scriptscriptfont1=\eightMath
\textfont2=\fourteenSymbols\scriptfont2=\tenSymbols\scriptscriptfont2=\eightSymbols
\textfont3=\fourteenMoreSymbols\scriptfont3=\tenMoreSymbols\scriptscriptfont3=\eightMoreSymbols
\textfont\itfam=\fourteenSansSerifItalics\def\it{\fam\itfam\fourteenSansSerifItalics}%
\textfont\slfam=\fourteenSansSerifSlanted\def\sl{\fam\slfam\fourteenSerifSansSlanted}%
\textfont\ttfam=\fourteenTypewriter\def\tt{\fam\ttfam\fourteenTypewriter}%
\textfont\bffam=\fourteenSansSerif%
\def\bf{\fam\bffam\fourteenSansSerif}\scriptfont\bffam=\tenSansSerif\scriptscriptfont\bffam=\eightSansSerif%
\def\cal{\fourteenSymbols}%
\def\greekbold{\fourteenMathBold}%
\def\gothic{\fourteenGothic}%
\def\Bbb{\fourteenDouble}%
\def\LieFont{\fourteenSerifItalics}%
\nt\normalbaselines\baselineskip=15pt%
}

\def\ChapterStyle{\parindent=0pt\parskip=0pt\normalbaselineskip=15pt%
\def\nt{\fourteenSansSerifBold}%
\def\rm{\fam0\fourteenSansSerifBold}%
\textfont0=\fourteenSansSerifBold\scriptfont0=\tenSansSerifBold\scriptscriptfont0=\eightSansSerifBold
\textfont1=\fourteenMath\scriptfont1=\tenMath\scriptscriptfont1=\eightMath
\textfont2=\fourteenSymbols\scriptfont2=\tenSymbols\scriptscriptfont2=\eightSymbols
\textfont3=\fourteenMoreSymbols\scriptfont3=\tenMoreSymbols\scriptscriptfont3=\eightMoreSymbols
\textfont\itfam=\fourteenSansSerifItalics\def\it{\fam\itfam\fourteenSansSerifItalics}%
\textfont\slfam=\fourteenSansSerifSlanted\def\sl{\fam\slfam\fourteenSerifSansSlanted}%
\textfont\ttfam=\fourteenTypewriter\def\tt{\fam\ttfam\fourteenTypewriter}%
\textfont\bffam=\fourteenSansSerif%
\def\bf{\fam\bffam\fourteenSansSerif}\scriptfont\bffam=\tenSansSerif\scriptscriptfont\bffam=\eightSansSerif%
\def\cal{\fourteenSymbols}%
\def\greekbold{\fourteenMathBold}%
\def\gothic{\fourteenGothic}%
\def\Bbb{\fourteenDouble}%
\def\LieFont{\fourteenSerifItalics}%
\nt\normalbaselines\baselineskip=15pt%
}

\def\SectionStyle{\parindent=0pt\parskip=0pt\normalbaselineskip=13pt%
\def\nt{\twelveSansSerifBold}%
\def\rm{\fam0\twelveSansSerifBold}%
\textfont0=\twelveSansSerifBold\scriptfont0=\eightSansSerifBold\scriptscriptfont0=\eightSansSerifBold
\textfont1=\twelveMath\scriptfont1=\eightMath\scriptscriptfont1=\eightMath
\textfont2=\twelveSymbols\scriptfont2=\eightSymbols\scriptscriptfont2=\eightSymbols
\textfont3=\twelveMoreSymbols\scriptfont3=\eightMoreSymbols\scriptscriptfont3=\eightMoreSymbols
\textfont\itfam=\twelveSansSerifItalics\def\it{\fam\itfam\twelveSansSerifItalics}%
\textfont\slfam=\twelveSansSerifSlanted\def\sl{\fam\slfam\twelveSerifSansSlanted}%
\textfont\ttfam=\twelveTypewriter\def\tt{\fam\ttfam\twelveTypewriter}%
\textfont\bffam=\twelveSansSerif%
\def\bf{\fam\bffam\twelveSansSerif}\scriptfont\bffam=\eightSansSerif\scriptscriptfont\bffam=\eightSansSerif%
\def\cal{\twelveSymbols}%
\def\bg{\twelveMathBold}%
\def\gothic{\twelveGothic}%
\def\Bbb{\twelveDouble}%
\def\LieFont{\twelveSerifItalics}%
\nt\normalbaselines\baselineskip=13pt%
}

\def\SubSectionStyle{\parindent=0pt\parskip=0pt\normalbaselineskip=13pt%
\def\nt{\twelveSansSerifItalics}%
\def\rm{\fam0\twelveSansSerifItalics}%
\textfont0=\twelveSansSerifItalics\scriptfont0=\eightSansSerifItalics\scriptscriptfont0=\eightSansSerifItalics%
\textfont1=\twelveMath\scriptfont1=\eightMath\scriptscriptfont1=\eightMath%
\textfont2=\twelveSymbols\scriptfont2=\eightSymbols\scriptscriptfont2=\eightSymbols%
\textfont3=\twelveMoreSymbols\scriptfont3=\eightMoreSymbols\scriptscriptfont3=\eightMoreSymbols%
\textfont\itfam=\twelveSansSerif\def\it{\fam\itfam\twelveSansSerif}%
\textfont\slfam=\twelveSansSerifSlanted\def\sl{\fam\slfam\twelveSerifSansSlanted}%
\textfont\ttfam=\twelveTypewriter\def\tt{\fam\ttfam\twelveTypewriter}%
\textfont\bffam=\twelveSansSerifBold%
\def\bf{\fam\bffam\twelveSansSerifBold}\scriptfont\bffam=\eightSansSerifBold\scriptscriptfont\bffam=\eightSansSerifBold%
\def\cal{\twelveSymbols}%
\def\greekbold{\twelveMathBold}%
\def\gothic{\twelveGothic}%
\def\Bbb{\twelveDouble}%
\def\LieFont{\twelveSerifItalics}%
\nt\normalbaselines\baselineskip=13pt%
}

\def\AuthorStyle{\parindent=0pt\parskip=0pt\normalbaselineskip=14pt%
\def\nt{\tenSerif}%
\def\rm{\fam0\tenSerif}%
\textfont0=\tenSerif\scriptfont0=\sevenSerif\scriptscriptfont0=\fiveSerif
\textfont1=\tenMath\scriptfont1=\sevenMath\scriptscriptfont1=\fiveMath
\textfont2=\tenSymbols\scriptfont2=\sevenSymbols\scriptscriptfont2=\fiveSymbols
\textfont3=\tenMoreSymbols\scriptfont3=\sevenMoreSymbols\scriptscriptfont3=\fiveMoreSymbols
\textfont\itfam=\tenSerifItalics\def\it{\fam\itfam\tenSerifItalics}%
\textfont\slfam=\tenSerifSlanted\def\sl{\fam\slfam\tenSerifSlanted}%
\textfont\ttfam=\tenTypewriter\def\tt{\fam\ttfam\tenTypewriter}%
\textfont\bffam=\tenSerifBold%
\def\bf{\fam\bffam\tenSerifBold}\scriptfont\bffam=\sevenSerifBold\scriptscriptfont\bffam=\fiveSerifBold%
\def\cal{\tenSymbols}%
\def\greekbold{\tenMathBold}%
\def\gothic{\tenGothic}%
\def\Bbb{\tenDouble}%
\def\LieFont{\tenSerifItalics}%
\nt\normalbaselines\baselineskip=14pt%
}

\def\AddressStyle{\parindent=0pt\parskip=0pt\normalbaselineskip=14pt%
\def\nt{\eightSerif}%
\def\rm{\fam0\eightSerif}%
\textfont0=\eightSerif\scriptfont0=\sevenSerif\scriptscriptfont0=\fiveSerif
\textfont1=\eightMath\scriptfont1=\sevenMath\scriptscriptfont1=\fiveMath
\textfont2=\eightSymbols\scriptfont2=\sevenSymbols\scriptscriptfont2=\fiveSymbols
\textfont3=\eightMoreSymbols\scriptfont3=\sevenMoreSymbols\scriptscriptfont3=\fiveMoreSymbols
\textfont\itfam=\eightSerifItalics\def\it{\fam\itfam\eightSerifItalics}%
\textfont\slfam=\eightSerifSlanted\def\sl{\fam\slfam\eightSerifSlanted}%
\textfont\ttfam=\eightTypewriter\def\tt{\fam\ttfam\eightTypewriter}%
\textfont\bffam=\eightSerifBold%
\def\bf{\fam\bffam\eightSerifBold}\scriptfont\bffam=\sevenSerifBold\scriptscriptfont\bffam=\fiveSerifBold%
\def\cal{\eightSymbols}%
\def\greekbold{\eightMathBold}%
\def\gothic{\eightGothic}%
\def\Bbb{\eightDouble}%
\def\LieFont{\eightSerifItalics}%
\nt\normalbaselines\baselineskip=14pt%
}

\def\AbstractStyle{\parindent=0pt\parskip=0pt\normalbaselineskip=12pt%
\def\nt{\eightSerif}%
\def\rm{\fam0\eightSerif}%
\textfont0=\eightSerif\scriptfont0=\sevenSerif\scriptscriptfont0=\fiveSerif
\textfont1=\eightMath\scriptfont1=\sevenMath\scriptscriptfont1=\fiveMath
\textfont2=\eightSymbols\scriptfont2=\sevenSymbols\scriptscriptfont2=\fiveSymbols
\textfont3=\eightMoreSymbols\scriptfont3=\sevenMoreSymbols\scriptscriptfont3=\fiveMoreSymbols
\textfont\itfam=\eightSerifItalics\def\it{\fam\itfam\eightSerifItalics}%
\textfont\slfam=\eightSerifSlanted\def\sl{\fam\slfam\eightSerifSlanted}%
\textfont\ttfam=\eightTypewriter\def\tt{\fam\ttfam\eightTypewriter}%
\textfont\bffam=\eightSerifBold%
\def\bf{\fam\bffam\eightSerifBold}\scriptfont\bffam=\sevenSerifBold\scriptscriptfont\bffam=\fiveSerifBold%
\def\cal{\eightSymbols}%
\def\greekbold{\eightMathBold}%
\def\gothic{\eightGothic}%
\def\Bbb{\eightDouble}%
\def\LieFont{\eightSerifItalics}%
\nt\normalbaselines\baselineskip=12pt%
}

\def\RefsStyle{\parindent=0pt\parskip=0pt%
\def\nt{\eightSerif}%
\def\rm{\fam0\eightSerif}%
\textfont0=\eightSerif\scriptfont0=\sevenSerif\scriptscriptfont0=\fiveSerif
\textfont1=\eightMath\scriptfont1=\sevenMath\scriptscriptfont1=\fiveMath
\textfont2=\eightSymbols\scriptfont2=\sevenSymbols\scriptscriptfont2=\fiveSymbols
\textfont3=\eightMoreSymbols\scriptfont3=\sevenMoreSymbols\scriptscriptfont3=\fiveMoreSymbols
\textfont\itfam=\eightSerifItalics\def\it{\fam\itfam\eightSerifItalics}%
\textfont\slfam=\eightSerifSlanted\def\sl{\fam\slfam\eightSerifSlanted}%
\textfont\ttfam=\eightTypewriter\def\tt{\fam\ttfam\eightTypewriter}%
\textfont\bffam=\eightSerifBold%
\def\bf{\fam\bffam\eightSerifBold}\scriptfont\bffam=\sevenSerifBold\scriptscriptfont\bffam=\fiveSerifBold%
\def\cal{\eightSymbols}%
\def\greekbold{\eightMathBold}%
\def\gothic{\eightGothic}%
\def\Bbb{\eightDouble}%
\def\LieFont{\eightSerifItalics}%
\nt\normalbaselines\baselineskip=10pt%
}

\def\ClaimStyle{\parindent=5pt\parskip=3pt\normalbaselineskip=14pt%
\def\nt{\tenSerifSlanted}%
\def\rm{\fam0\tenSerifSlanted}%
\textfont0=\tenSerifSlanted\scriptfont0=\sevenSerifSlanted\scriptscriptfont0=\fiveSerifSlanted
\textfont1=\tenMath\scriptfont1=\sevenMath\scriptscriptfont1=\fiveMath
\textfont2=\tenSymbols\scriptfont2=\sevenSymbols\scriptscriptfont2=\fiveSymbols
\textfont3=\tenMoreSymbols\scriptfont3=\sevenMoreSymbols\scriptscriptfont3=\fiveMoreSymbols
\textfont\itfam=\tenSerifItalics\def\it{\fam\itfam\tenSerifItalics}%
\textfont\slfam=\tenSerif\def\sl{\fam\slfam\tenSerif}%
\textfont\ttfam=\tenTypewriter\def\tt{\fam\ttfam\tenTypewriter}%
\textfont\bffam=\tenSerifBold%
\def\bf{\fam\bffam\tenSerifBold}\scriptfont\bffam=\sevenSerifBold\scriptscriptfont\bffam=\fiveSerifBold%
\def\cal{\tenSymbols}%
\def\greekbold{\tenMathBold}%
\def\gothic{\tenGothic}%
\def\Bbb{\tenDouble}%
\def\LieFont{\tenSerifItalics}%
\nt\normalbaselines\baselineskip=14pt%
}

\def\ProofStyle{\parindent=5pt\parskip=3pt\normalbaselineskip=14pt%
\def\nt{\tenSerifSlanted}%
\def\rm{\fam0\tenSerifSlanted}%
\textfont0=\tenSerif\scriptfont0=\sevenSerif\scriptscriptfont0=\fiveSerif
\textfont1=\tenMath\scriptfont1=\sevenMath\scriptscriptfont1=\fiveMath
\textfont2=\tenSymbols\scriptfont2=\sevenSymbols\scriptscriptfont2=\fiveSymbols
\textfont3=\tenMoreSymbols\scriptfont3=\sevenMoreSymbols\scriptscriptfont3=\fiveMoreSymbols
\textfont\itfam=\tenSerifItalics\def\it{\fam\itfam\tenSerifItalics}%
\textfont\slfam=\tenSerif\def\sl{\fam\slfam\tenSerif}%
\textfont\ttfam=\tenTypewriter\def\tt{\fam\ttfam\tenTypewriter}%
\textfont\bffam=\tenSerifBold%
\def\bf{\fam\bffam\tenSerifBold}\scriptfont\bffam=\sevenSerifBold\scriptscriptfont\bffam=\fiveSerifBold%
\def\cal{\tenSymbols}%
\def\greekbold{\tenMathBold}%
\def\gothic{\tenGothic}%
\def\Bbb{\tenDouble}%
\def\LieFont{\tenSerifItalics}%
\nt\normalbaselines\baselineskip=14pt%
}

%
%


\def\ModeYes{yes}
\def\ModeNo{no}

\def\ModeUndef{undefined}


\def\nx{\noexpand}
\def\ni{\noindent}
\def\newpage{\vfill\eject}

\def\ss{\vskip 5pt}
\def\ms{\vskip 10pt}
\def\bs{\vskip 20pt}

 \def\,{\mskip\thinmuskip}
 \def\!{\mskip-\thinmuskip}
 \def\>{\mskip\medmuskip}
 \def\;{\mskip\thickmuskip}

%
% ... 
%

\def\refsModePost{post}
\def\refsModeAuto{auto}

\def\dbRefsSatusModeOk{ok}
\def\dbRefsSatusModeError{error}
\def\dbRefsSatusModeWarning{warning}


\newcount\BNUM
\BNUM=0

\def\refs{}

\def\SetModePost{\xdef\refsMode{\refsModePost}}			
\SetModePost

\def\dbRefsStatusOk{%
	\xdef\dbRefsStatus{\dbRefsSatusModeOk}%
	\xdef\dbRefsError{\ModeNo}%
	\xdef\dbRefsWarning{\ModeNo}%
	\xdef\dbRefsInfo{\ModeNo}%
}

\def\dbRefs{%
}

\def\dbRefsGet#1{%
	\xdef\found{N}\xdef\ikey{#1}\dbRefsStatusOk%
	\xdef\key{\ModeUndef}\xdef\tag{\ModeUndef}\xdef\tail{\ModeUndef}%
	\dbRefs%
}

\def\NextRefsTag{%
	\global\advance\BNUM by 1%
}
\def\ShowTag#1{{\bf [#1]}}

\def\dbRefsInsert#1#2{%
\dbRefsGet{#1}%
\if\found Y %
   \xdef\dbRefsStatus{\dbRefsSatusModeWarning}%
   \xdef\dbRefsWarning{record is already there}%
   \xdef\dbRefsInfo{record not inserted}%
\else%
   \toks2=\expandafter{\dbRefs}%
   \ifx\refsMode\refsModeAuto \NextRefsTag
    \xdef\dbRefs{%
   	\the\toks2 \nx\xdef\nx\dbx{#1}%
	\nx\ifx\nx\ikey %
		\nx\dbx\nx\xdef\nx\found{Y}%
		\nx\xdef\nx\key{#1}%
		\nx\xdef\nx\tag{\the\BNUM}%
		\nx\xdef\nx\tail{#2}%
	\nx\fi}%
	\global\xdef\refs{\refs \ss\ni[\the\BNUM]\ #2\par}
   \fi%
   \ifx\refsMode\refsModePost 
    \xdef\dbRefs{%
   	\the\toks2 \nx\xdef\nx\dbx{#1}%
	\nx\ifx\nx\ikey %
		\nx\dbx\nx\xdef\nx\found{Y}%
		\nx\xdef\nx\key{#1}%
		\nx\xdef\nx\tag{\ModeUndef}%
		\nx\xdef\nx\tail{#2}%
	\nx\fi}%
   \fi%
\fi%
}

\def\dbRefsEdit#1#2#3{\dbRefsGet{#1}%
\if\found N 
   \xdef\dbRefsStatus{\dbRefsSatusModeError}%
   \xdef\dbRefsError{record is not there}%
   \xdef\dbRefsInfo{record not edited}%
\else%
   \toks2=\expandafter{\dbRefs}%
   \xdef\dbRefs{\the\toks2%
   \nx\xdef\nx\dbx{#1}%
   \nx\ifx\nx\ikey\nx\dbx %
	\nx\xdef\nx\found{Y}%
	\nx\xdef\nx\key{#1}%
	\nx\xdef\nx\tag{#2}%
	\nx\xdef\nx\tail{#3}%
   \nx\fi}%
\fi%
}

\def\bib#1#2{\RefsStyle\dbRefsInsert{#1}{#2}%
	\ifx\dbRefsStatus\dbRefsSatusModeWarning %
		\message{^^J}%
		\message{WARNING: Reference [#1] is doubled.^^J}%
	\fi%
}

\def\ref#1{\dbRefsGet{#1}%
\ifx\found N %
  \message{^^J}%
  \message{ERROR: Reference [#1] unknown.^^J}%
  \ShowTag{??}%
\else%
	\ifx\tag\ModeUndef \NextRefsTag%
		\dbRefsEdit{#1}{\the\BNUM}{\tail}%
		\dbRefsGet{#1}%
		\global\xdef\refs{\refs \ss\ni [\tag]\ \tail\par}
	\fi
	\ShowTag{\tag}%
\fi%
}

\def\ShowBiblio{\ms\Ensure{\SectionEnsure}%
{\SectionStyle\ni References}%
{\RefsStyle\refs}%
}

\newcount\CHANGES
\CHANGES=0
\def\AuxFile{7}
\def\PreventDoubleOn{\xdef\PreventDoubleLabel{\ModeYes}}

\PreventDoubleOn

\def\StoreLabel#1#2{\xdef\itag{#2}
 \ifx\PreModeStatus\ModeNo %
   \message{^^J}%
   \errmessage{You can't use Check without starting with OpenPreMode (and finishing with ClosePreMode)^^J}%
 \else%
   \immediate\write\AuxFile{\nx\dbLabelPreInsert{#1}{\itag}}%
   \dbLabelGet{#1}%
   \ifx\itag\tag %
   \else%
	\global\advance\CHANGES by 1%
 	\xdef\itag{(?.??)}%
    \fi%
   \fi%
}

\def\PreModeStatus{\ModeNo}

\def\ClosePreMode{\immediate\closeout\AuxFile%
  \ifnum\CHANGES=0%
	\message{^^J}%
	\message{**********************************^^J}%
	\message{**  NO CHANGES TO THE AuxFile  **^^J}%
	\message{**********************************^^J}%
 \else%
	\message{^^J}%
	\message{**************************************************^^J}%
	\message{**  PLAEASE TYPESET IT AGAIN (\the\CHANGES)  **^^J}%
    \errmessage{**************************************************^^ J}%
  \fi%
  \edef\PreModeStatus{\ModeNo}%
}

\def\dbLabelSatusModeOk{ok}

\def\dbLabelSatusModeWarning{warning}

\def\dbLabelStatusOk{%
	\xdef\dbLabelStatus{\dbLabelSatusModeOk}%
	\xdef\dbLabelError{\ModeNo}%
	\xdef\dbLabelWarning{\ModeNo}%
	\xdef\dbLabelInfo{\ModeNo}%
}

\def\dbLabel{%
}

\def\dbLabelGet#1{%
	\xdef\found{N}\xdef\ikey{#1}\dbLabelStatusOk%
	\xdef\key{\ModeUndef}\xdef\tag{\ModeUndef}\xdef\pre{\ModeUndef}%
	\dbLabel%
}

\def\ShowLabel#1{%
 \dbLabelGet{#1}%
 \ifx\tag \ModeUndef %
 	\global\advance\CHANGES by 1%
 	(?.??)%
 \else%
 	\tag%
 \fi%
}

\def\dbLabelPreInsert#1#2{\dbLabelGet{#1}%
\if\found Y %
  \xdef\dbLabelStatus{\dbLabelSatusModeWarning}%
   \xdef\dbLabelWarning{Label is already there}%
   \xdef\dbLabelInfo{Label not inserted}%
   \message{^^J}%
   \errmessage{Double pre definition of label [#1]^^J}%
\else%
   \toks2=\expandafter{\dbLabel}%
    \xdef\dbLabel{%
   	\the\toks2 \nx\xdef\nx\dbx{#1}%
	\nx\ifx\nx\ikey %
		\nx\dbx\nx\xdef\nx\found{Y}%
		\nx\xdef\nx\key{#1}%
		\nx\xdef\nx\tag{#2}%
		\nx\xdef\nx\pre{\ModeYes}%
	\nx\fi}%
\fi%
}

\def\dbLabelInsert#1#2{\dbLabelGet{#1}%
\xdef\itag{#2}%
\dbLabelGet{#1}%
\if\found Y %
	\ifx\tag\itag %
	\else%
	   \ifx\PreventDoubleLabel\ModeYes %
		\message{^^J}%
		\errmessage{Double definition of label [#1]^^J}%
	   \else%
		\message{^^J}%
		\message{Double definition of label [#1]^^J}%
	   \fi%
	\fi%
   \xdef\dbLabelStatus{\dbLabelSatusModeWarning}%
   \xdef\dbLabelWarning{Label is already there}%
   \xdef\dbLabelInfo{Label not inserted}%
\else%
   \toks2=\expandafter{\dbLabel}%
    \xdef\dbLabel{%
   	\the\toks2 \nx\xdef\nx\dbx{#1}%
	\nx\ifx\nx\ikey %
		\nx\dbx\nx\xdef\nx\found{Y}%
		\nx\xdef\nx\key{#1}%
		\nx\xdef\nx\tag{#2}%
		\nx\xdef\nx\pre{\ModeNo}%
	\nx\fi}%
\fi%
}


\newcount\PART
\newcount\CHAPTER
\newcount\SECTION
\newcount\SUBSECTION
\newcount\FNUMBER

\PART=0
\CHAPTER=0
\SECTION=0
\SUBSECTION=0	
\FNUMBER=0

\def\LastPart{\ModeUndef}
\def\LastChapter{\ModeUndef}
\def\LastSection{\ModeUndef}
\def\LastSubSection{\ModeUndef}
\def\LastClaim{\ModeUndef}
\def\Last{\ModeUndef}

\newdimen\TOBOTTOM
\newdimen\LIMIT

\def\Ensure#1{\ \par\ \immediate\LIMIT=#1\immediate\TOBOTTOM=\the\pagegoal\advance\TOBOTTOM by -\pagetotal%
\ifdim\TOBOTTOM<\LIMIT\newpage \else%
\vskip-\parskip\vskip-\parskip\vskip-\baselineskip\fi}

\def\PartLabel{\the\PART}
\def\NewPart#1{\global\advance\PART by 1%
         \bs\ni{\PartStyle  Part \PartLabel:}
         \bs\ni{\PartStyle #1}\newpage%
         \CHAPTER=0\SECTION=0\SUBSECTION=0\FNUMBER=0%
         \gdef\Left{#1}%
         \global\edef\Last{\PartLabel}%
         \global\edef\LastPart{\PartLabel}%
         \global\edef\LastChapter{\ModeUndef}%
         \global\edef\LastSection{\ModeUndef}%
         \global\edef\LastSubSection{\ModeUndef}%
         \global\edef\LastClaim{\ModeUndef}}
\def\ChapterLabel{\the\CHAPTER}
\def\NewChapter#1{\global\advance\CHAPTER by 1%
         \bs\ni{\ChapterStyle  Chapter \ChapterLabel: #1}\ms%
         \SECTION=0\SUBSECTION=0\FNUMBER=0%
         \gdef\Left{#1}%
         \global\edef\Last{\ChapterLabel}%
         \global\edef\LastChapter{\ChapterLabel}%
         \global\edef\LastSection{\ModeUndef}%
         \global\edef\LastSubSection{\ModeUndef}%
         \global\edef\LastClaim{\ModeUndef}}
\def\SectionEnsure{3cm}
\def\NewSection#1{\Ensure{\SectionEnsure}\gdef\SectionLabel{\the\SECTION}\global\advance\SECTION by 1%
         \ms\ni{\SectionStyle  \SectionLabel.\ #1}\ss%
         \SUBSECTION=0\FNUMBER=0%
         \gdef\Left{#1}%
         \global\edef\Last{\SectionLabel}%
         \global\edef\LastSection{\SectionLabel}%
         \global\edef\LastSubSection{\ModeUndef}%
         \global\edef\LastClaim{\ModeUndef}}
\def\NewAppendix#1#2{\Ensure{\SectionEnsure}\gdef\SectionLabel{#1}\global\advance\SECTION by 1%
         \bs\ni{\SectionStyle  Appendix \SectionLabel.\ #2}\ss%
         \SUBSECTION=0\FNUMBER=0%
         \gdef\Left{#2}%
         \global\edef\Last{\SectionLabel}%
         \global\edef\LastSection{\SectionLabel}%
         \global\edef\LastSubSection{\ModeUndef}%
         \global\edef\LastClaim{\ModeUndef}}
\def\Acknowledgements{\Ensure{\SectionEnsure}\gdef\SectionLabel{}%
         \ms\ni{\SectionStyle  Acknowledgments}\ss%
         \SECTION=0\SUBSECTION=0\FNUMBER=0%
         \gdef\Left{}%
         \global\edef\Last{\ModeUndef}%
         \global\edef\LastSection{\ModeUndef}%
         \global\edef\LastSubSection{\ModeUndef}%
         \global\edef\LastClaim{\ModeUndef}}
\def\SubSectionEnsure{2cm}
\def\SubSectionLabel{\ifnum\SECTION>0 \the\SECTION.\fi\the\SUBSECTION}
\def\NewSubSection#1{\Ensure{\SubSectionEnsure}\global\advance\SUBSECTION by 1%
         \ms\ni{\SubSectionStyle #1}\ss%
         \global\edef\Last{\SubSectionLabel}%
         \global\edef\LastSubSection{\SubSectionLabel}}
\def\SetNumberingModeN{\def\ClaimLabel{(\the\FNUMBER)}}
\def\SetNumberingModeSN{\def\ClaimLabel{(\ifnum\SECTION>0 \SectionLabel.\fi%
      \the\FNUMBER)}}
\def\SetNumberingModeCSN{\def\ClaimLabel{(\ifnum\CHAPTER>0 \the\CHAPTER.\fi%
      \ifnum\SECTION>0 \SectionLabel.\fi%
      \the\FNUMBER)}}

\def\NewClaim{\global\advance\FNUMBER by 1%
    \ClaimLabel%
    \global\edef\LastClaim{\ClaimLabel}%
    \global\edef\Last{\ClaimLabel}}

\def\HideLabels{\xdef\ShowLabelsMode{\ModeNo}}
\HideLabels

\def\fn{\eqno{\NewClaim}} 
\def\fl#1{%
\ifx\ShowLabelsMode\ModeYes%
 \eqno{{\buildrel{\hbox{\AbstractStyle[#1]}}\over{\hfill\NewClaim}}}%
\else%
 \eqno{\NewClaim}%
\fi%
\dbLabelInsert{#1}{\ClaimLabel}}
\def\fprel#1{\global\advance\FNUMBER by 1\StoreLabel{#1}{\ClaimLabel}%
\ifx\ShowLabelsMode\ModeYes%
\eqno{{\buildrel{\hbox{\AbstractStyle[#1]}}\over{\hfill.\itag}}}%
\else%
 \eqno{\itag}%
\fi%
}

\def\cl#1{\global\advance\FNUMBER by 1\dbLabelInsert{#1}{\ClaimLabel}%
\ifx\ShowLabelsMode\ModeYes%
${\buildrel{\hbox{\AbstractStyle[#1]}}\over{\hfill\ClaimLabel}}$%
\else%
  $\ClaimLabel$%
\fi%
}
\def\cprel#1{\global\advance\FNUMBER by 1\StoreLabel{#1}{\ClaimLabel}%
\ifx\ShowLabelsMode\ModeYes%
${\buildrel{\hbox{\AbstractStyle[#1]}}\over{\hfill.\itag}}$%
\else%
  $\itag$%
\fi%
}

\def\Note{\ms\leftskip 3cm\rightskip 1.5cm\AbstractStyle}
\def\endNote{\par\leftskip 2cm\rightskip 0cm\NormalStyle\ss}


\parindent=7pt
\leftskip=2cm
\newcount\SideIndent
\newcount\SideIndentTemp
\SideIndent=0
\newdimen\SectionIndent
\SectionIndent=-8pt

\def\sidebar{\vrule height15pt width.2pt }
\def\endcorner{\hbox{\hbox{\vrule height6pt width.2pt}\vbox to6pt{\vfill\hbox
to4pt{\leaders\hrule height0.2pt\hfill}}}}
\def\begincorner{\hbox{\hbox{\vrule height6pt width.2pt}\vbox to6pt{\hbox
to4pt{\leaders\hrule height0.2pt\hfill}}}}
\def\endbegincorner{\hbox{\vbox to15pt{\endcorner\vskip-6pt\begincorner\vfill}}}
\def\SideShow{\SideIndentTemp=\SideIndent \ifnum \SideIndentTemp>0 
\loop\sidebar\hskip 2pt \advance\SideIndentTemp by-1\ifnum \SideIndentTemp>1 \repeat\fi}

\def\BeginSection{{\vbadness 100000 \par\ni\hskip\SectionIndent%
\SideShow\vbox to 15pt{\vfill\begincorner}}\global\advance\SideIndent by1\vskip-10pt}

\def\EndSection{{\vbadness 100000 \par\ni\global\advance\SideIndent by-1%
\hskip\SectionIndent\SideShow\vbox to15pt{\endcorner\vfill}\vskip-10pt}}

\def\EndBeginSection{{\vbadness 100000\par\ni%
\global\advance\SideIndent by-1\hskip\SectionIndent\SideShow
\vbox to15pt{\vfill\endbegincorner}}%
\global\advance\SideIndent by1\vskip-10pt}

\def\ShowBeginCorners#1{%
\SideIndentTemp =#1 \advance\SideIndentTemp by-1%
\ifnum \SideIndentTemp>0 %
\vskip-15truept\hbox{\kern 2truept\vbox{\hbox{\begincorner}%
\ShowBeginCorners{\SideIndentTemp}\vskip-3truept}}%
\fi%
}

\def\ShowEndCorners#1{%
\SideIndentTemp =#1 \advance\SideIndentTemp by-1%
\ifnum \SideIndentTemp>0 %
\vskip-15truept\hbox{\kern 2truept\vbox{\hbox{\endcorner}%
\ShowEndCorners{\SideIndentTemp}\vskip 2truept}}%
\fi%
}

\def\BeginSections#1{{\vbadness 100000 \par\ni\hskip\SectionIndent%
\SideShow\vbox to 15pt{\vfill\ShowBeginCorners{#1}}}\global\advance\SideIndent by#1\vskip-10pt}

\def\EndSections#1{{\vbadness 100000 \par\ni\global\advance\SideIndent by-#1%
\hskip\SectionIndent\SideShow\vbox to15pt{\vskip15pt\ShowEndCorners{#1}\vfill}\vskip-10pt}}

\def\EndBeginSections#1#2{{\vbadness 100000\par\ni%
\global\advance\SideIndent by-#1%
\hbox{\hskip\SectionIndent\SideShow\kern-2pt%
\vbox to15pt{\vskip15pt\ShowEndCorners{#1}\vskip4pt\ShowBeginCorners{#2}}}}%
\global\advance\SideIndent by#2\vskip-10pt}




%
%


\def\al{\alpha}

\def\om{\omega}
\def\si{\sigma}
\def\vp{\varphi}

\def\Si{\Sigma}


 \def\calS{{\hbox{\cal S}}}




 \def\R{{\hbox{\Bbb R}}}

 \def\Z{{\hbox{\Bbb Z}}}

 \def\R{{\hbox{\Bbb R}}}


\def\Aut{{\hbox{Aut}}}

\def\Diff{{\hbox{Diff}}}

\def\id{{\hbox{\rm id}}}

\def\ip{\hbox to4pt{\leaders\hrule height0.3pt\hfill}\vbox to8pt{\leaders\vrule width0.3pt\vfill}\kern 2pt}
\def\QDE{\hfill\hbox{\ }\vrule height4pt width4pt depth0pt} 
\def\del{\partial}

\def\arr{\rightarrow}

%
%

\def\DEFINITION{\ClaimStyle\ni{\bf Definition: }}

\def\ENDDEFINITION{\NormalStyle}

\def\LEMMA{\ClaimStyle\ni{\bf Lemma: }}

\def\ENDLEMMA{\NormalStyle}

\def\THEOREM{\ClaimStyle\ni{\bf Theorem: }}

\def\ENDTHEOREM{\NormalStyle}

\def\PROOF{\ProofStyle\ni{\bf Proof: }}
\def\ENDPROOF{\hfill\QDE\NormalStyle}

\def\cases#1{\left\{\eqalign{#1}\right.}
\NormalStyle
\SetNumberingModeSN
\PreventDoubleOn

\long\def\title#1{\centerline{\TitleStyle\ni#1}}

\long\def\author#1{\ms\centerline{\AuthorStyle by {\it #1}}}

\long\def\address#1{\ss\centerline{\AddressStyle #1}\par}
\long\def\moreaddress#1{\centerline{\AddressStyle #1}\par}
\def\abstract{\ms\leftskip 3cm\rightskip .5cm\AbstractStyle{\bf \ni Abstract:}\ }
\def\endabstract{\par\leftskip 2cm\rightskip 0cm\NormalStyle\ss}

\SetNumberingModeSN

\def\calG{{\hbox{\cal G}}}

\def\calD{{\hbox{\cal D}}}

\def\strictsubset{\hbox{\Bbb\char32}}

\def\frac[#1/#2]{\hbox{$#1\over#2$}}
\def\Frac[#1/#2]{{#1\over#2}}
\def\({\left(}
\def\){\right)}
\def\[{\left[}
\def\]{\right]}

\def\^#1{{}^{#1}_{\>\cdot}}
\def\_#1{{}_{#1}^{\>\cdot}}
\def\Label=#1{{\buildrel {\hbox{\fiveSerif \ShowLabel{#1}}}\over =}}
\def\<{\kern -1pt}


\def\ExpandAllCNotes{\long\def\CNote##1{%
\BeginSection
	\Note%
 		##1%
	\endNote%
\EndSection%
}}
\ExpandAllCNotes
%
%
%
%


\def\frame#1{\vbox{\hrule\hbox{\vrule\vbox{\kern2pt\hbox{\kern2pt#1\kern2pt}\kern2pt}\vrule}\hrule\kern-4pt}}

\def\Box to #1#2#3{\frame{\vtop{\hbox to #1{\hfill #2 \hfill}\hbox to #1{\hfill #3 \hfill}}}}


\bib{HawkingEllis}{S.W.Hawking,  G.Ellis,
{\it The Large Scale Structure of the Universe},
 Cambridge Univ. Press (1973)
}

\bib{Einstein}{A.Einstein,
{\it Die formale Grundlage der allgemeinen Relativitts-theorie}
Kniglich Preussische Akademie der Wissenschaften (Berlin), Sitzungsberichte (1914) 1030--1085
}

\bib{Wald}{R.Wald,
{\it General Relativity},
 (Chicago, IL, University of Chicago Res (1984)
}

\bib{Norton1}{J.D.Norton,
{\it General covariance and the foundations of general relativity: eight decades of dispute},
Rep.Prog.Phys. {\bf 56} (1993) 791-858
}

\bib{Norton2}{J.D.Norton,
{\it General Covariance, Gauge Theories and the Kretschmann Objection}
In:Symmetries in physics, Brading, K. (ed.) et al.:  (2003) 110-123}

\bib{Stackel}{M. Iftime, J.Stackel,
{\it The Hole Argument for Covariant Theories},
Gen.Rel.Grav. 38 (2006) 1241--1252
}

\bib{Geroch}{R.Geroch, 
{\it Gauge, Diffeomorphisms, Initial-Value Formulation, etc The Einstein equations and the large scale behavior of gravitational fields}, 
Birkhuser, Basel, (2004) 441--477
}

\bib{Simon1}{S. Garruto, L.Fatibene,
{\it The Cauchy Problem in General Relativity: an Algebraic Characterization},
(in preparation)
}

\bib{Rovelli1}{C.Rovelli,  
{\it Halfway through the Woods: Contemporary Research on Space and Time},
 in J. Earman and J. D. Norton (eds.) The Cosmos of Science: Essays of Exploration. Pittsburgh: University of Pittsburgh Press (1997) 180-223
}

\bib{Rovelli2}{C.Rovelli, 
{\it What is observable in classical and quantum gravity?},
Class. Quantum Grav. {\bf 8}(3) (1991) 297--316
}

\bib{Godesics}{L.Fatibene, M.Francaviglia, G. Magnano, 
{\it On a Characterization of Geodesic Trajectories and Gravitational Motions}, 
Int. J. Geom. Methods Mod. Phys.9(5) (2012) 1220007
}

\bib{Bosons}{L. Fatibene, M. Ferraris, M. Francaviglia,
{\it Noether formalism for conserved quantities in classical gauge 
field theories. II. The arbitrary bosonic matter case},
J. Math. Phys. {\bf 38}(8) (1997) 3953--3967
}

\bib{Spinors}{L. Fatibene, M. Ferraris, M Francaviglia, M. Godina,
{\it Gauge formalism for general relativity and fermionic matter},
Gen. Relativity Gravitation 30(9) (1998), 1371--1389
}

\bib{Book}{L. Fatibene, M. Francaviglia,
{\it Natural and gauge natural formalism for classical field theories. A geometric perspective including spinors and gauge theories},
Kluwer Academic Publishers, Dordrecht, 2003. xxii+365 pp. ISBN: 1-4020-1703-0}

\bib{Kretschmann}{E.Kretschmann,
{\it \"Uber den physikalischen Sinn der Relativit\"atspostulate, A.Einsteins neue und seine urspr\"ungliche Relativit\"atstheorie},
Annalen der Physik, 53 (1917)
575Ð614}

\bib{Giulini}{D.Giulini,
{\it Some remarks on the notions of general covariance and background independence},
Lect. Notes Phys. 721 (2007) 105--120}

\bib{Anderson}{J.L.Anderson,
{\it Recent Developments in General Relativity},
 ed. Editorial Commitee (New York MacMillan) (1964) Gravitation and relativity 
 eds. H Y Chiu and W F Hoffmann (New York.Benjamin)}

\bib{GNStructures}{L Fatibene, M Ferraris, M Francaviglia,
{\it On the gauge natural structure of modern physics}
Int. J. Geom. Methods Mod. Phys., 1(4), 443 (2004). DOI: 10.1142/S0219887804000253}

\bib{Kepler}{L.Fatibene, M.Francaviglia, S. Mercadante,
{\it Mathematical Equivalence vs. Physical Equivalence between Extended Theories of Gravitations},
Int.J.Geom.Meth.Mod.Phys. 11 (2014) 1450008; arXiv:1302.2938 [gr-qc]
}

\bib{Janiska}{J. Janyska,
{\it Higher order Utiyama invariant interaction}  
In: Rep. Math. Phys. 59 (2007), 63--81}

\bib{Utiyama}{R.Utiyama,
{\it Invariant theoretical interpretation of interaction}
Phys. Rev. 101 (1956), 1597--1607}

\bib{Weinberg}{S. Weinberg, 
{\it Gravitation and Cosmology: Principles and Applications of the General Theory of Relativity}, 
Wiley, New York (a.o.) (1972). XXVIII, 657 S. : graph. Darst.. ISBN: 0-471-92567-5}

\bib{HoleLusanna}{Luca Lusanna, Massimo Pauri,
{\it Explaining Leibniz equivalence as difference of non-inertial appearances: Dis-solution of the Hole Argument and physical individuation of point-events},
Studies in History and Philosophy of Science Part B: Studies in History and Philosophy of Modern Physics, Volume 37, Issue 4, December 2006, Pages 692?725}

\bib{ADM}{R.  Arnowitt, S.  Deser and C.  W.  Misner, in: 
{\it Gravitation: An Introduction to Current Research}, 
L. Witten ed. Wyley,  227, (New York, 1962); gr-qc/0405109}

\bib{Nester}{J. Nester, J. Isenberg,
{\it Canonical Gravity},
in: {\it General Relativity and Gravitation. One Hundred Years After the Birth of Albert Einstein. Vol. 1}
A. Held eds. Plenum Press NY (1980)}

\bib{Lusanna}{L.Lusanna,
{\it Canonical ADM tetrad gravity: From metrological inertial gauge variables to dynamical tidal Dirac observables},
Int. J. Geom. Met. Mod. Phys., {\bf 12}(3), 2015 1530001; arXiv:1108.3224 [gr-qc]}

\bib{NostroADM}{L. Fatibene, M. Ferraris, M. Francaviglia, L. Lusanna, 
{\it ADM Pseudotensors, Conserved Quantities and Covariant Conservation Laws in General Relativity}, 
Annals of Physics {\bf 327}(6),  (2012),  1593?1616; arXiv:gr-qc/1007.4071}

\bib{Rezzolla}{L.Rezzolla, O.Zanotti, 
{\it Relativistic Hydrodynamics}, 
Oxford (2013) ISBN: 9780198528906}

\bib{3+1}{E. Gourgoulhon.
{\it 3+1 Formalism and Bases of Numerical Relativity}; arXiv:gr-qc/0703035,  (2007)}

\bib{Simon2}{L.Fatibene, S. Garruto, 
{\it Principal Symbol of Euler-Lagrange Operators}; arXiv:1603.04732 [gr-qc]}



\def\ubal{\underline{\al}\kern1pt}
\def\obal{\overline{\al}\kern1pt}

\def\ubR{\underline{R}\kern1pt}
\def\obR{\overline{R}\kern1pt}
\def\ubom{\underline{\om}\kern1pt}
\def\obxi{\overline{\xi}\kern1pt}
\def\ubu{\underline{u}\kern1pt}
\def\ube{\underline{e}\kern1pt}
\def\obe{\overline{e}\kern1pt}

\NormalStyle

\title{Constraining the Physical State by Symmetries}

\author{L.Fatibene$^{a,b}$, M.Ferraris$^{a}$, G.Magnano$^{a}$, }

\address{$^a$ Department of Mathematics, University of Torino (Italy)}
\moreaddress{$^b$ INFN - Sezione Torino - IS QGSKY}

\abstract
After reviewing the hole argument and its relations with initial value problem and general covariance,
we shall discuss how much freedom one has to define the physical state of a system in a generally covariant (or gauge covariant) field theory. 

We shall show that in gauge covariant theories (and generally covariant theories with a a compact space) one has no freedom and one is forced to declare as physically equivalent
two configurations which differ by a gauge transformation (or by a global spacetime diffeomorphism), as it is usually prescribed.

On the contrary, when space is not compact, the result proven for the compact case does not hold true and one may have different options to define physically equivalent configurations,
still preserving determinism. 
\endabstract

\NewSection{Introduction}

After almost a century since formulation of General Relativity (GR) it is still not clear and unanimously accepted what exactly Einstein discovered and what are the foundations of GR; see  \ref{Norton1} and references quoted therein.
The emphasis on different assumptions (covariance principle in its active or passive form, equivalence principle in its weak or strong form, Mach's principle, coincidence principle just to quote the most used ones) has been moved many times since 1915 and moved differently by different authors. Not an exception is the meaning of the hole argument (which will be reviewed below) and the problem of observability, i.e.~the definition of what has to be understood by {\it physical state} in a covariant theory; see \ref{Rovelli2}. Originally, Einstein formulated the {\it hole argument} in terms of boundary problems (see \ref{Einstein})
while soon after Hilbert reformulated it in terms of Cauchy problems.

Back in 1917 Kretschmann raised an argument claiming that (just because of Einstein coincidence principle) any theory can be reformulated into a covariant form and as a consequence covariance principle has no physical content; see \ref{Kretschmann}.
As a matter of fact neither Kretschmann argument nor coincidence principle has ever been given a precise form. 
In particular although his argument seems convincing and well supported by  relevant examples, it was never clarified  what one should understand by {\it any theory} or which transformations one should be ready to apply to a theory to make it generally covariant.
Since then, the consensus for Kretschmann has grown from an isolated critique to mainstream and the covariance principle has been progressively marginalised to a heuristic role. 
On the other hand, the hole argument can be used also with gauge theories and gauge covariance, while apparently nobody claims that gauge covariance is a vacuous principle; see \ref{GNStructures}.

At the same time, Utiyama-like arguments (see \ref{Janiska}, \ref{Bosons}, \ref{Spinors}) have been routinely used to in fact select covariant dynamics, i.e.~to do exactly what Kretschmann argument says covariance principles should be unable to do. Moreover, the equivalence principle, which has gradually replaced covariance principle as the accepted physical core of GR (see \ref{Weinberg}, \ref{Wald}), has been shown to be a consequence of essentially covariance requests (at least in its weak form, i.e.~as the claim that the free fall motion of any body is described by geodesics equations); see \ref{Godesics}.

Norton described this situation and noticed that one is not even forced to take a stand for one or the other party; see \ref{Norton2}.
In Kretschmann's argument he was in fact discussing the possibility of setting up a covariant formulation for any theory, possibly at the price of introducing as many fields and equations as needed with whatever meaning is needed to achieve the goal of covariantisation, which seems to be a reasonable claim.
In all other arguments (from Utiyama to weak equivalence principle) one instead works with a fixed set of fields, in a Lagrangian setting or with additional requirements such as absence of background structures;
see \ref{Anderson}, \ref{Giulini}.

There is also another issue strictly connected to the hole argument.
In most cases (actually, to the best of our knowledge, in all of them) the hole argument is formulated in terms of compactly supported diffeomorphisms; see \ref{Weinberg}, \ref{Wald}, \ref{Stackel}, \ref{HoleLusanna}. The name of the argument refers to a compact region, the {\it hole}, in which the diffeomorphism is not the identity.
Originally Einstein argued that by a diffeomorphism supported in the hole and in view of active general covariance, one could show that the value of the metric field in the region is not determined by its value outside the region and on its boundary. 

However, by analysing the hole argument is terms of initial value problems, the holes emerge naturally to be much more general. 
What is needed is in fact that one can find a Cauchy surface out of the hole, not requiring in general the hole to be compact at all; see \ref{Simon1}.  
In fact one can consider diffeomorphisms which are the identity in a region around a Cauchy surface enlarging the hole to cover the whole $M$ except a neighbourhood of the Cauchy surface. These spacetime diffeomorphisms will be called {\it Cauchy-compatible transformations} or {\it Cauchy transformations} for short.
A Cauchy transformation leaves a Cauchy surface untouched with a neighbourhood of it, i.e.~it preserves the initial conditions. On the other hand, it produces a different solution since it is not the identity away from the Cauchy surface, compromising the uniqueness of solutions in Cauchy problems.
 
This form of the hole argument for Cauchy problems shows that essentially general covariance cannot mathematically coexist with Cauchy well--posed problems.
Norton also argued (see \ref{Norton1}) that the definition of physical state is something to be discussed in physics and it is not something which can be settled by a purely mathematical argument.
This position is certainly plausible and agreeable. 
However, we shall here argue that {\it some} constraints to the definition of physical state can be in fact put on a mathematical stance (the ones which go back to Einstein-Hilbert about well-poseness of Cauchy problem or boundary problem).  
And a detailed analysis of these issues shows an unexpected structure of cases which is not clarified in general, yet.
What is clear is that assuming, as usually done, that  the physical state of a covariant theory is to be identified with equivalence classes of configurations modulo spacetime diffeomorphisms
is a fair assumption, still a choice which is sometimes forced by mathematics (in particular by determinism in the form of Cauchy theorem) but sometimes it is one of many possible choices which,  in those cases, we agree should be addressed from a physical stance.

The relation between general (and gauge) covariance and observables in a theory has been analysed in detail in the literature; see \ref{Norton1}, \ref{Norton2}, \ref{Giulini}, \ref{Stackel}, \ref{Rovelli1}.
Einstein general covariance, his hole argument and Kretschmann's counterargument have been discussed since 1917; see \ref{Kretschmann}.

In most cases general covariance has been identified in its active form with covariance with respect to general diffeomorphisms in spacetime and accordingly two configurations which are mapped one into the other by a spacetime diffeomorphism are considered to represent the same single physical state.
We shall here take a more minimalistic attitude and discuss what are the mathematical constraint in defining physical states.
We shall discuss that sometimes one can find subclasses of diffeomorphisms (the group generated by Cauchy-compatible transformations)
which play a distinctive role in the discussion and which to the best of our knowledge have not properly been taken into account in standard frameworks.
 
\ms

Let us start, for the sake of simplicity,  by restricting to generally covariant theories. Gauge theories will be briefly discussed in the conclusions since most of what we shall do easily applies to those more general cases, as well; see \ref{Book} for general framework and notation.
 
In a generally covariant theory one has a huge group of symmetries $\calS$ containing the (lift to the configuration bundle of the) spacetime diffeomorphisms $\Diff(M)$, and in particular
spacetime diffeomorphisms which can be connected by a flow with the identity $\id_M$, namely $\Diff_e(M)$.  
Any element $\Phi\in \Diff_e(M)$ can be obtained by evaluating a 1-parameter subgroup $\Phi_s$ at $s=1$, i.e.~$\Phi=\Phi_1$. The 1-parameter subgroup $\Phi_s$ is also called a {\it flow} of diffeomorphisms. 

The standard attitude is to assume that in a generally covariant theory  any two configurations of fields differing by {\it any spacetime diffeomorphism} represent the same physical state.
In other words, if $\si$ is a section of the configuration bundle and $\Phi_\ast \si=\si'$ is its image through a diffeomorphism (in $\Diff_e(M)$ or in $\Diff(M)$ depending on the case)
then both $\si$ and $\si'$ represent the same physical state of the system.

Let us call {\it gauge transformations} the group of transformations $\calG$ which fix the physical state. 
As a matter of fact defining the group of gauge transformations is equivalent to define the physical state. 
Either one defines the physical state as the orbits of the action of gauge transformations
or defines gauge transformations as the transformations mapping one representative of the physical state into another representative of the same physical state.

In order for this to make sense one needs a gauge transformation $\Phi$ to be a symmetry of the system (as it is in generally covariant theories) since if $\si$ is a solution, of course also $\si'$ must be a solution as well. In other words any gauge transformation (acting on configurations but leaving the physical state unchanged) must be a symmetry and the symmetry group is a upper bound to the group of gauge transformations, i.e.~one must have $\calG\subset \calS$.
 
We shall show that there is  a lower bound (which will be denoted by $\vec \calD$)  for the gauge group $\calG$  as well, i.e.~one must have $\vec\calD\subset \calG\subset \calS$. 

We shall argue that in principle the group $\vec\calD\subset \calS=\Diff_e(M)$ leaves a certain freedom in setting 
$\calG$ between its lower and upper bounds, i.e.~that one may have $\vec\calD\>\strictsubset\> \calS$. 
In these cases one has different options to set  the gauge group $\vec\calD\subset \calG\subset \calS$ and
each different assumption about what $\calG$ is  in fact defines a different theory with the same dynamics but different interpretation of what is the physical state and what can be in principle observed.
We shall also discuss topological conditions on $M$ for which this freedom is nullified and one is forced to set $\calG= \Diff_e(M)$ as usually done in the literature.

\Note
If you consider Kepler mechanical system, the symmetry group $\calS$ contains rotations (and in that case one can show that $\vec\calD=\{e\}$ is trivial). 
In that situations one is free to consider rotations as gauge transformations or not. That corresponds to considering absolute angles as observables or not. 
That in turn goes back to decide if the observer has access to fixed stars (to be used as a reference to measure absolute angles) or not (so that measuring absolute angles becomes impossible). The two options can be considered as two different physical systems since we can measure different things in the two cases.
\endNote

The lower bound for $\calG$ is set by requiring field equations to be deterministic on the physical states by implementing the {hole argument}. 
Let us  first review this argument in its essential form and directly connect it to determinism of field equations.
The same argument is the standard proof that field equations in a (general or gauge) covariant  theory are not deterministic for the configurations, i.e.~the configuration are a redundant  description of the physical state, which is the essence of gauge theories.

\NewSubSection{ADM splittings and Cauchy problems}

In order to present the hole argument we have to state Cauchy theorem for field equations. For that we need to introduce {\it ADM splittings} of $M$.

\ms
\DEFINITION
An ADM splitting of a spacetime $M$ is a bundle structure $t: M\arr \R$. 
\ENDDEFINITION 
\ss
The projection $t$ associates to any event $x\in M$ a number $t(x)=t_0\in \R$ which called the ADM-time for $x$.
The fibers $t^{-1}(t_0)\subset M$ are called {\it isochronous hypersurfaces} (at the time $t_0$), they are denoted also by $t^{-1}(t_0)= \Si_{t_0}$ and they are diffeomorphic to the standard fiber $\Si \simeq \Si_{t_0}$. 
Since the base $\R$ of the ADM splitting is contractible, the bundle is necessarily trivial, i.e. one has global trivializations and $M\simeq \R\times \Si$. 
Accordingly, a global ADM splitting is not possible on general manifolds. If a spacetime manifold allows an ADM splitting it must be diffeomorphic to a product $\R\times \Si$ and it is called a
{\it globally hyperbolic} manifold.  In order to be able to define a well-posed Cauchy problem we have to restrict the spacetime $M$ to be globally hyperbolic.
Hence hereafter we shall assume $M\simeq \R\times \Si$.

The fields $\si$ can be decomposed to fields on the space hypersurface $\Si$. 
In a Cauchy problem one assigns the value of fields and their (time)-derivatives on a Cauchy surface $\Si_{0}$ and field equations determine their evolution in time and allow 
to rebuild the covariant field $\si$ which is a solution of field equations; see \ref{ADM}, \ref{Nester}, \ref{Lusanna}, \ref{NostroADM}, \ref{Rezzolla}, \ref{3+1}, \ref{Simon1}, \ref{Simon2}.

\Note
To be true some of field equations only give constraints on initial conditions and, at the same time, some of the spatial fields on $\Si$ are left undetermined and parametrize the ADM splittings. 
That is another story not completely unrelated to symmetries but let us overview this viewpoint.
\endNote

\NewSubSection{The hole argument}

Let us consider a spacetime diffeomorphism $\Phi:M\arr M$ which is the identity on an open set  $t^{-1}(I)$ for some open interval $I\subset \R$ (and it is different from the identity somewhere).
Two configurations $\si$ and $\Phi_\ast \si$ have the same initial conditions on a Cauchy surface $\Si_0$ which projects on a time $t_0\in I\subset \R$.
Such a diffeomorphism $\Phi$ which is the identity around a Cauchy surface will be called a {\it Cauchy transformation}. 
The set of all Cauchy diffeomorphisms will be denoted by $\calD^1\subset \Diff_e(M)$.
Since $\Phi\in \calD^1$ is also a symmetry, if $\si$ is a solution then $\si'=\Phi_\ast \si$ is also a solution, a {\it different} solution corresponding to the very same initial conditions set on $\Si_0$.
Accordingly, initial conditions and field equations do not uniquely select a solution and the Cauchy problem is not well-posed.

The only way out if one wants to save determinism at least for the physical states  is to identify $\si$ and $\si'$ as two different representations of one single physical state.
One can say that there is an equivalence relation $\sim$ which defines the physical states as equivalence classes of configurations. 
As far as different representatives of the same physical state are connected by symmetries, field equations are compatible to quotient induced by the equivalence relation $\sim$
and define field equations on physical states which are deterministic.

The group of gauge transformations $\calG$ should contain at least finite compositions of Cauchy diffeomorphisms.
We can set $\calD^2$ for diffeomorphisms which can be written as a composition of two Cauchy diffeomorphisms, i.e.
$$
\calD^2=\{ \Phi=\phi_1\circ \phi_2: \phi_1, \phi_2\in \calD^1\}
\fn$$
Of course, one has $\calD^1\subset \calD^2$ and, in general, neither $\calD^1$ or $\calD^2$ are groups since they are not closed by composition.

Thus one can iterate the definition of $\calD^n$ for diffeomorphisms which can be written as a composition of $n$ Cauchy diffeomorphisms
and $\vec \calD$ for diffeomorphisms which can be written as a composition of a finite number of Cauchy diffeomorphisms.
Then one has the inclusions
$$
\calD^1\subset \calD^2\subset \dots \subset \calD^n\subset \dots \subset \vec \calD\subset \Diff_e(M)\equiv \calS
\fn$$
The set $\vec \calD\subset \Diff_e(M)$ is a group and it is the lower bound for gauge transformations. If one did not have $\vec \calD\subset \calG$, then the relation $\sim$ defined by
gauge equivalence would not induce deterministic field equations for the physical states.
Thus one can set gauge transformations at will, provided that one has $\vec \calD\subset \calG \subset \calS$.

Now in some cases it may happen that $\vec \calD = \calS$ so that one is forced to set $\calG= \calS$ as usually done in relativistic and gauge theories.
We shall show below that for example this is the case every time $\Si$ is compact. 
As a special case one has the relativistic point particles for which $\Si$ is compact since it is made of  one point only and one is forced to accept that two curves in spacetime which differ
by a reparametrization do describe the same physical state. This is why parametrizations of worldlines are depleted of a  direct physical meaning and two different parametrizations of the same worldline trajectory do describe the same motion.  

\Note
Let us stress the argument just presented above once again. We have general mathematical constraints for what should be the physical state for a relativistic particle.
One can show that these constraints imply that reparametrizations of the worldline trajectories are gauge transformations.
Then we assume that the physical state is described by worldline trajectories.
This is purely a mathematical argument.

On the other hand we can discuss the motion in spacetimes of particles on a physical stance and show that it is reasonable to assume that the physical state is described by worldline trajectories and parametrizations are irrelevant. The two viewpoints come (quite independently) to the same conclusion, which is a good thing.

We shall show that this is not the general situation. There exist cases in which the mathematical constraints do not force a definition but leave some (though not complete) freedom which has to be fixed by further physical considerations. 
However, until mathematics forces the choice or puts constraints to what should be physically decided we believe it is worth doing it on a mathematical stance.
\endNote

We shall also show a counter example, showing  a globally hyperbolic spacetime $M=\R\times \R$ in which the situation is different from the compact space case.  
As a consequence the usual assumption of identifying configurations which differ by a diffeomorphism is a legitimate though in general unmotivated choice.
When describing a system one should be aware of which assumptions come from mathematical constraints and which assumptions are done on a physical stance.

When setting up a covariant theory one should first study whether the group $\vec\calD$ is a strict subgroup of $\Diff_e(M)$. If it is,  one should characterise possible subgroups
$\vec \calD\> \strictsubset\> \calG\>\strictsubset\> \calS$. Then one should declare which of such groups $\vec \calD\> \strictsubset\> \calG\>\strictsubset\> \calS$ is elected as a gauge group.
Different choices lead to different theories with an equivalent dynamics but different observables.

\ 

\NewSection{Compact Space Case}

Hereafter we shall prove that if the space manifold $\Si$ is compact then necessarily one has $\calD^1\> \strictsubset\> \calD^2=\dots=\vec \calD  = \calS$.
In these theories one has no freedom and has to set $\calG=\Diff_e(M)$. Two configurations which are mapped one into the other by a spacetime diffeomorphism  $\Phi\in\Diff_e(M)$ (has it is usually done) define the same physical state.

Let us consider an ADM splitting of $M\simeq \R\times \Si$ with $\Si$ a compact $(m-1)$-dimensional manifold.

\ms
\LEMMA
If the transformation $\Phi=\Phi_{s=1}\in \Diff_e(M)$ can be written as a composition of two Cauchy diffeomorphisms $\Phi=\phi_1\circ \phi_2$  (with $\phi_1,\phi_2\in \calD^1$)
then the diffeomorphism $\hat \Phi:=\Phi\circ \phi^{-1}_2= \phi_1$ is the identity in a region $t^{-1}(I_1)\subset M$ for some open set $I_1\subset \R$ and it coincides with the given $\Phi$
  in a region $t^{-1}(I_2)\subset M$ for some open set $I_2\subset \R$.
\ENDLEMMA
\ss

\PROOF
In fact both $\phi_1$ and $\phi_2$ are Cauchy transformations and hence they coincide with the identity in $t^{-1}(I_1)$ and $t^{-1}(I_2)$, respectively.
In the region $t^{-1}(I_1)$ one has that $\phi_1$ is the identity and $\hat\Phi=\phi_1$ coincides with the identity as well.
In the region  $t^{-1}(I_2)$ one has that $\phi_2$ is the identity and $\hat\Phi=\Phi$ coincides with the original transformation $\Phi$.
\ENDPROOF
\ms

Such a diffeomorphism $\hat \Phi$ which is the identity in a region $t^{-1}(I_1)\subset M$ for some open set $I_1\subset \R$ and it coincides with the given $\Phi$
in a region $t^{-1}(I_2)\subset M$ for some open set $I_2\subset \R$ is called a {\it splitter} for $\Phi$.

Thus we can look for a splitter $\hat\Phi$ for $\Phi$.
If we find it then we can write
$$
\Phi= \(\Phi \circ \hat \Phi^{-1}\) \circ \hat\Phi
$$ 
and $\hat\Phi$ is a Cauchy transformation since it is the identity in the region $t^{-1}(I_1)$ while $\Phi \circ \hat \Phi^{-1}$ is a Cauchy transformation as it is the identity in the region $t^{-1}(I_2)$.
Thus $\Phi\in \calD^2$.
Accordingly, the diffeomorphism $\Phi\in \Diff_e(M)$ is in $\calD^2$ iff there exists a splitter for it.

\Note
First let us notice that we can define a function $\hat\vp:\R\arr \R$ supported in $[-\al, \al]$, smooth everywhere, with a maximum in $0$ such that 
$\hat\vp(0)=1$.
For example we can set 
$$
\hat\vp(t)=\cases{
 \exp\(- \frac[a^2t^2/\al^2-t^2]\) \quad& t\in (-\al, \al)\cr
 0\qquad\qquad\qquad& \hbox{otherwise}
 }
\fn$$
Thus the function 
$$
\vp_+(t)=\Frac[1/N]\int^t_{-\infty} \hat\vp(t') dt' 
\fn$$
and fix $N(a)$ so that $\vp_+(\al)=1$.
The function $\vp_+(t)$ is a step-like smooth functions which is 0 in the region $t\le -\al$ and $1$ in the region $t\ge \al$ with the derivative in 
$[-\al, \al]$ which is in  $(0,\frac[1/\al]]$.

Let us also define the function $\vp_-(t)=1-\vp_+(t)$ which is decreasing, is 1 in the region $t\le -\al$ and $0$ in the region $t\ge \al$ with the derivative in  $[-\al, \al]$ which is in  $[-\frac[1/\al], 0)$.

\endNote

Let us show that:
\ms
\THEOREM
For any $\Phi\in \Diff_e(M)$ we can define a splitter $\phi$ for $\Phi$.
\ENDTHEOREM
\ss
\PROOF
Since  $\Phi\in \Diff_e(M)$ we have a flow $\Phi_s$ such that $\Phi=\Phi_{(s=1)}$.
The infinitesimal generator of that flow $\xi$ is a complete spacetime vector field.

Then let us consider the vector field $\xi_+= \vp_+(t(x))\cdot \xi$, denote its flow by $\hat\phi_{s}$ and set $\phi=\hat\phi_{(s=1)}$
which is a diffeomorphism by construction.

Now let us notice that an integral curve of $\xi_+$ is a reparametrization of an integral curve of $\xi$.
If we set initial condition in the region $t(x)>\al$ (where $\hat\xi=\xi$) and we have to follow the flow $\hat\phi_{s}$ until $s=1$, this may not be as following the flow $\Phi_s$ up to $s=1$. In fact, at some point, before $s=1$, the integral curve may get to the region $t(x)<\al$, feeling a generator $\xi_+\not=\xi$ and slow down with respect to the integral curve of $\Phi_s$.

In order to avoid this, one should consider curves that will not exit from the region $t(x)>\al$. This can easily be done by considering the space 
$\Si_{\al}$ at $t=\al$ and considering $\Si'=\Phi^{-1}(\Si_{\al})$: an integral curve starting at a point on the right of $\Si'$ will be an integral curve of 
$\xi$ at least until the parameter is $s\le 1$.

Now consider the function $f:\Si\arr \R: x\mapsto t(\Phi^{-1}(x))$. This is defined on a compact and it attends its maximum, say $t_0$.
Then if one considers the integral curve of $\xi_+$ starting at a point $x_0$ with $t(x_0)>t_0$, 
then such a curve is an integral curve of $\xi$ as well and,
accordingly, one has $\hat\phi(x_0)= \Phi(x_0)$.

Then the two diffeomorphisms $\hat \phi$ and $\Phi$ coincide in the region $t(x)>t_0$ and $\hat \phi=\id_M$ in the region $t(x)<-\al$. 
Hence in any event the map $\phi$ is a splitter for $\Phi$ and then $\Phi\in\calD^2$.
\ENDPROOF
\ms
Since any $\Phi\in \Diff_e(M)$ is also in $\Phi\in\calD^2$ then one has
$$
\calD^1\>\strictsubset\>\calD^2= \dots = \calD^n= \vec \calD=\Diff_e(M)
\fn$$
and one has no option than to set $\calG= \Diff_e(M)$.
All diffeomorphisms are gauge transformations and they preserve the physical state.

In other words, when $\Si$ is compact, the usual choice of setting $\calG=\Diff_e(M)$ is forced (and supported) by a purely mathematical argument.

\NewSection{The spatial non-compact case}
The argument for compact spaces does not work in the non-compact case since one is unable to guarantee that the function $f$ has a maximum.
It would be quite natural to conjecture that also for spatially non-compact case one still had $\calD^2= \Diff_e(M)$, possibly by a different proof.

However, this is false in general. Let us first exhibit a counterexample.
Let us consider $M=\R^2$ which is split by the ADM foliation as $M\simeq \R\times \R\arr \R$; ADM coordinates on $\R^2$ are $(t, x)$ and the projection reads as $\pi:\R^2\arr \R:(t, x)\mapsto t$.
Let us consider the diffeomorphism $\Phi:\R^2\arr \R^2\in \Diff_e(M)$ given by
$$
\cases{
&t'= \cos(t^2+x^2) t - \sin(t^2+x^2)x\cr
&x'=\- \sin(t^2+x^2)t+ \cos(t^2+x^2) x\cr
}
\fn$$
which is in fact a diffeomorphism since its Jacobian is nowhere degenerate (being $J=1$ everywhere) and being in the flow of the vector field
$\xi=(t^2+x^2)(x\del_t-t\del_x)$.
One can easily check that $\Phi\not\in \calD^1$ since its fixed points are the origin and a discrete sets of circles around it (the ones of radius $r^2=t^2+x^2= 2k\pi$ with $k\in \Z$)
so that $\Phi$ is not the identity on any vertical line.

The map $\Phi$ maps vertical stripes around in a spiral as shown in Figure $1$. 
Thus the spiral is driven around and forced to have points on the right and on the left of any other vertical stripe.
Accordingly, if for some reason one has that the image of a vertical stripe needs to be all on one side of some other vertical stripe a contradiction is obtained.
In other words, one has precisely that the function $f$ defined for the compact space case above attends no maximum or minimum.

\ms

\centerline{
   \includegraphics[width=8cm]{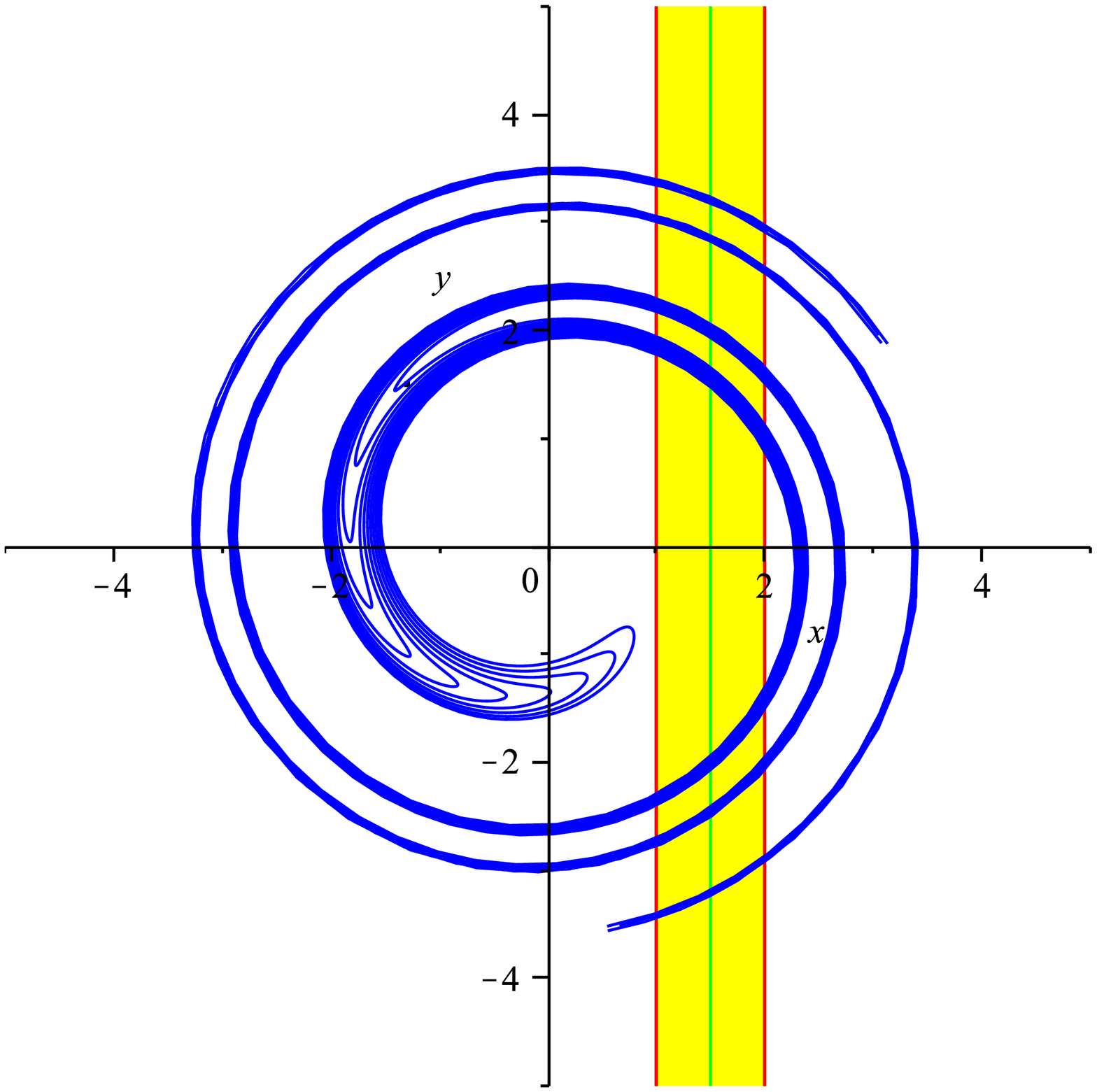} 
}
\centerline{\it
   { Figure 1: Mapping a vertical stripe (light) into a spiral (dark) }}
\centerline{\it
   { by means of the diffeomorphism $\Phi$}
}

\ms

One can easily show that $\Phi\not\in \calD^2$; for, let us suppose for the sake of argument that one can write $\Phi=\phi_1\circ \phi_2$
with $\phi_1,\phi_2\in \calD^1$ and denote by $U_i= \pi^{-1}(I_i)$ with $I_i$ intervals in $\R$ so that $\phi_i$ is the identity on the vertical stripe $U_i$.
The two intervals $I_1$ and $I_2$ are disjoint (if not $\Phi$ would be in $\calD^1$).
With no loss of generality we can choose $I_1 < I_2$ (meaning that any point in $I_1$ is on the left of any point in $I_2$);
if this were not the case one can repeat the argument for $\Phi^{-1}= \phi_2^{-1}\circ \phi_1^{-1}$ (being that $\Phi\in \calD^2$ if and only if $\Phi^{-1}\in \calD^2$).

In view of the projection $\pi$ and the ordering of the base manifold $\R$, one can define a partial ordering of points in $\R^2$ writing $p<q$ if $\pi(p)<\pi(q)$.
As in $\R$, one can write $p<U$ for a subset $U\subset \R^2$ if $\forall q\in U: p<q$. 
Finally, given two subsets $U, U'\subset \R^2$ one can write $U<U'$ if $\forall p\in U$ and $q\in U': p<q$.
Let us remark that a map $\phi_1\in \calD^1$ which is the identity on a stripe $U_1$ maps the left (right) region into itself.
Thus one has that for any $p<U_1$ then one has $\phi_1(p)<U_1$.

Now consider a point $p\in U_2$; one has $\phi_2(p)= p$ and $\phi_1(p)=\Phi(p)$. Since $\phi_1$ is the identity on $U_1$ and $U_1<p$ then $U_1< \Phi(p)$.
Thus the diffeomorphism $\Phi$ maps the stripe $U_2$ to the right of $U_1$, i.e.~we write $U_1<\Phi(U_2)$.
However, since the stripe $U_2$ go up and down to infinity, the spiral $\Phi(U_2)$ is driven around and sooner or later it will be forced to be on the left of $U_1$.
By the contradiction one concludes that $\Phi$ cannot the written as $\Phi=\phi_1\circ \phi_2$, i.e.~$\Phi\not\in \calD^2$.

\ms
Thus in the cases for which the space manifold is non-compact one can have diffeomorphisms of $M$ which are not in $\calD^2$ and, accordingly, 
one has $\calD^2\> \strictsubset\> \Diff_e(M)$.

With some more effort one can show that the diffeomorphism $\Phi$ here considered is not in $\calD^3$; see Appendix $A$ below.
Thus one has that $\calD^3\> \strictsubset\> \Diff_e(M)$.
However, this direct approach to $\calD^n$ grows too difficult since this direct proof needs to consider in $\calD^n$ all the possible orderings of the stripes $U_i$
on which the diffeomorphisms $\phi_i$ reduce to the identity once one assumes that $\Phi=\phi_1\circ\phi_2\circ\dots \circ \phi_n$.
In other words the length of the direct proof grows as $n!/2$ and it becomes soon too long.
One can improve the situation (though not substantially) by considering some general results; see Appendix $B$ below. However, further investigation in that direction is needed.

\ms

To the best of our knowledge it is still unclear whether one has $\vec \calD =\Diff_e(M)$ in general or in the spatial non-compact case one could have 
$\vec \calD \> \strictsubset\> \Diff_e(M)$.
We stress that this problem should be addressed {\it before} defining the physical state, for example by setting $\calG=\Diff_e(M)$.
In fact, if one had $\vec \calD \> \strictsubset\> \Diff_e(M)$ all one could know for sure is that the gauge group should be chosen to be
$\vec \calD \subset \calG \subset \Diff_e(M)$ and at least two different options (namely, $\vec \calD = \calG$ and $\calG =\Diff_e(M)$) exist.

Different choices correspond to different classes of configurations representing the physical states, and hence quantities which are observable by one choice
may be unobservable with the other. When this freedom exists, the definition of the physical state has to be decided on a physical ground 
and different choices correspond to different definitions of the system.
The standard choice $\calG =\Diff_e(M)$ is legitimate and it is in a sense canonical.
Still, one should know if it was a choice among many possibilities or it was forced by the nature of the problem, for example as it happens when the space manifold is compact.

\NewSection{Conclusions and Perspectives}

We showed that the group $\calG$ of gauge transformations needs to contain the group $\vec\calD$ generated by Cauchy diffeomorphisms
(otherwise initial conditions do not single out uniquely a solution for field equations)
and must be contained in the symmetry group $\calS=\Diff_e(M)$ of the theory 
(otherwise field equations are not compatible with the definition of physical state as equivalence class of configurations under the action of $\calG$).

In globally hyperbolic spacetimes $M\simeq \R\times \Si$ if $\Si$ is compact then one can prove $\vec\calD= \calS$ and one is forced to fix $\calG$
to the whole symmetry group.

When $\Si$ is non-compact then one still has $\vec\calD\subset \calS$ but in some cases the inclusion may be strict.
Hence in these cases one has some freedom and different options to fix the group  $\calG$ of gauge transformations.

If one now considers gauge covariant theories, the group of symmetries contains all flows of vertical principal automorphisms of the structure bundle $P$, i.e.~$\calS=\Aut_V(P)$.
Let us denote by $\pi$ the projection of the structure bundle $P$ and by $t:M\arr \R$ the ADM projection.
Such flows $\Phi_s$  are locally generated by pure gauge transformations, namely they are in the form
$$
\cases{
&x'^\mu=x^\mu\cr
&g'= \al_s(x)\cdot g\cr
}
\fn$$ 
for some local function $\al_s(x)\in G$.
As we can dump diffeomorphisms in a region, we can also dump vertical automorphisms by defining
$$
\hat \Phi_+ (p)= \Phi_{\vp(t(\pi(p))}(p)
\qquad
\hat \Phi_- (p)= \Phi_{1-\vp(t(\pi(p))}(p)
\fn$$
Both $\hat \Phi_\pm$ are Cauchy gauge transformations in $\calD^1$ and
$$
\eqalign{
\hat \Phi_+ \circ \hat \Phi_- (p)=&\Phi_{\vp(t(\pi(\Phi_{1-\vp(t(\pi(p))}(p)))}\circ \Phi_{1-\vp(t(\pi(p))}(p)
= \Phi_{\vp(t(\pi(p))}\circ \Phi_{1-\vp(t(\pi(p))}(p)=\cr
=&\Phi_{\vp(t(\pi(p))+1-\vp(t(\pi(p))}(p)
=\Phi_{1}(p)=\Phi(p)
}
\fn$$
Thus any $\Phi\in \Aut_V(P)$ is in fact in $\calD^2$ and $\calD^1 \subset \calD^2 = \dots =  \calD^n=\vec\calD=  \Aut_V(P)$.
 Again one has no freedom in choosing $\calG$ and it necessarily agrees with $ \calG=\Aut_V(P)=\calS$ confirming the assumptions which is generally done in the literature.

Thus the only case in which one can have the gauge group strictly smaller than the symmetry group is in general relativistic theories on a  spacetime with a non-compact space.
To the best of our knowledge, this problem seems not to have been considered in the literature where the standard choice $\calG=\calS$ is assumed without discussion.
On the other hand, further investigations are needed to possibly produce a clear counter example showing that one could have $\vec\calD\>\strictsubset \> \calS$
as we conjectured could happen (or to prove that the conjecture is false and the standard choice is forced in general on a mathematical stance).

\NewAppendix{A}{Results for $\calD^3$}

Let us assume notation as in Section $3$.
Let us now show that $\Phi\not\in \calD^3$. Again for the sake of arguments let us suppose that $\Phi=\phi_1\circ\phi_2\circ \phi_3$ and, with the same notation as above, assume that $I_1<I_3$.

Case (a): one has $I_1 < I_3< I_2 $. Then let us set  $J= \phi_3^{-1}(U_2)$. We have $U_3<J$ and $U_1<U_3<\phi_3(J)=U_2$. 
Accordingly, one should have $U_1<\Phi(J)= \phi_1(U_2)$ and $J$ is a stripe, not a vertical stripe but still a stripe going to infinity on both sides.
Thus, $\Phi$ drives the stripe $J$ around in the spiral and sooner or later one is forced to have points in $\Phi(J)$ on the left of $U_1$, which is a contradiction.

Case (b): one has $I_1 < I_2 < I_3$. Let us set $J=U_3$.
Then one has $U_2<\phi_3(U_3)=U_3$, and $U_1<U_2<\phi_2\circ \phi_3(U_3)$. Thus one should have $U_1<\Phi(U_3)$.
However, $\Phi$ drives the stripe around in the spiral and sooner or later one is forced to have points in $\Phi(U_3)$ on the left of $U_1$, which is a contradiction.

Case (c):  one has $I_2 < I_1 < I_3$. Consider $J= \phi_3^{-1}(U_2)< U_3$. Then  one has $\phi_3(J)=U_2<U_1$ and $\phi_2\circ \phi_3(J)=U_2<U_1$ and accordingly,
$\Phi(J)< U_1$. However, 
$\Phi$ drives the stripe around in the spiral and sooner or later one is forced to have points in $\Phi(J)$ on the right of $U_1$, which is a contradiction.

Since one cannot have any of these cases and $I_1, I_2, I_3$ are pairwise disjoint intervals (otherwise $\Phi$ would be in $\calD^2$)
then one can conclude that $\Phi\not\in \calD^3$.

A similar proof for $\calD^4$ should consider $4!/2= 12$ possible subcases of different orderings.

\ 

\NewAppendix{B}{General Results for $\calD^n$}

Actually, some general result is available and could simplify the general proof for $\calD^n$.

For example, in $\calD^3$ one can show that some of the orderings cannot appear since they actually correspond to morphisms in $\calD^2$.

For example,  let us assume $\Phi=\phi_1\circ \phi_2\circ \phi_3\in \calD^3$ with $I_1<I_2<I_3$ and where $\phi_i\in \calD^1$ are Cauchy diffeomorphisms.
Each of the Cauchy transformation $\phi_i$ can be split as $\phi_i= \phi_i^+\circ \phi_i^-=\phi_i^-\circ \phi_i^+$
where $\phi_i^\pm$ are the identity on the left (on the right) of $I_i$.

Then one has that $\phi_i^+\circ \phi_i^-=\phi_i^-\circ \phi_i^+$ and that for example
$\phi_1^+\circ \phi_2^+=\tilde \phi_1^+$ (and similarly $\phi_1^-\circ \phi_2^-=\tilde \phi_2^-$, respectively)
where the transformation $\tilde \phi_1^+$ ($\tilde \phi_2^-$, respectively) is the identity on the left (right, respectively) of $I_1$ ($I_2$, respectively).

Then one has a set of moves to simplify a sequence. For example:
$$
\eqalign{
\phi_1\circ \phi_2\circ \phi_3=& \phi_1^+\circ (\phi_1^-\circ \phi_2^-) \circ (\phi_2^+\circ \phi_3^+)\circ \phi_3^-=
\phi_1^+\circ \tilde \phi_2^- \circ \tilde \phi_2^+\circ \phi_3^-=\cr
=& \phi_1^+ \circ \tilde \phi_2^+\circ \tilde \phi_2^-\circ \phi_3^-
= \tilde \phi_1^+ \circ \tilde \phi_3^- \in \calD^2
}\fn$$

Analogously one can prove that a transformation $\Phi=\phi_1\circ \phi_2\circ \phi_3\in \calD^3$ with $I_3<I_2<I_1$ is in fact in $\calD^2$.
Accordingly, whenever one finds a sequence of three Cauchy transformations which are the identity on stripes $U_i$ which are ordered as (or in reverse order)
the composition ordering, they actually simplify to lower order.

This lemma simplifies  the proof that $\Phi\not\in \calD^4$.
For example let us consider a map $\Phi= \phi_1\circ \phi_2\circ \phi_3\circ \phi_4\in \calD^4$  with $I_3<I_4< I_2< I_1$, this is actually in $\calD^3$
since the stripes $I_3< I_2< I_1$ appears in reverse order.

By this techniques one can show that in fact only $5$ of the $12$ possible orderings need to be analysed for $\calD^4$.
On a similar basis one needs to analyse $14$ of the  $60$ possible orderings need to be analysed for $\calD^5$.
However, one would need to investigate if there exist other cases which simplify besides the ordered triples.

\Acknowledgements

We wish to thank A. Albano, A. Fino, E. Winterroth for useful discussions and comments. We also acknowledge the contribution of INFN (Iniziativa Specifica QGSKY), the local research project {\it  Metodi Geometrici in Fisica Matematica e Applicazioni (2015)} of Dipartimento di Matematica of University of Torino (Italy). This paper is also supported by INdAM-GNFM
and by the COST action CA15117 (CANTATA).

\ShowBiblio

\end